DOI: 10.1002/marc.201600281**Communication**

# Generation of silicone poly-HIPES with controlled pore sizes via reactive emulsion stabilization[a]

Anaïs Giustiniani\*, Philippe Guégan, Manon Marchand, Christophe Poulard and Wiebke Drenckhan

_________

A. Giustiniani, M. Marchand, Asso. Pr. C. Poulard, Dr. W. Drenckhan,
Laboratoire de Physique des Solides, CNRS, Univ. Paris-Sud, Université Paris-Saclay,
91405 Orsay Cedex, France
E-mail: giustiniani@lps.u-psud.fr; poulard@lps.u-psud.fr; drenckhan@lps.u-psud.fr
Dr. P. Guégan,
Institut Parisien de Chimie Moléculaire, Université Pierre et Marie Curie
3, rue Galilée, 94200 Ivry sur Seine (France)
E-mail: philippe.guegan@upmc.fr

_________Macrocellular silicone polymers are obtained after solidification of the continuous phase of a PDMS (polydimethylsiloxane) emulsion, which contains PEG (polyethylene glycol) drops of sub-millimetric dimensions. Coalescence of the liquid template emulsion is prohibited by a reactive blending approach. We investigate in detail the relationship between the interfacial properties and the emulsion stability, and we use micro- and millifluidic techniques to generation macro-cellular polymers with controlled structural properties over a wider range of cell-sizes (0.2-2mm) and volume fractions of the continuous phase (0.1-40%). This approach could easily be transferred to a wide range of polymeric systems.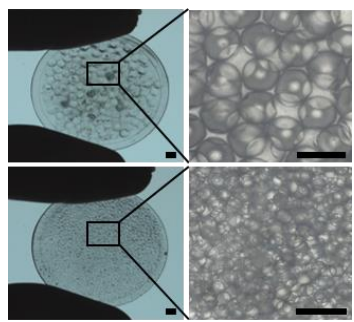

---

[a] **Supporting Information** is available online from the Wiley Online Library or from the author.

- 1 -

# 1. Introduction

Macro-cellular polymers are highly searched-for materials thanks to their rich physical properties. These arise from the internal structuration of the material, in which discrete cells of gas or liquid are tightly packed within a continuous polymeric solid. The size, shape, organisation and relative volume of these cells have an important influence on the overall material properties. Commonly, the cellular organisation is the signature of an initially liquid state, in which gas bubbles (foam) or liquid droplets (emulsion) are compacted within a continuous liquid monomer/polymer matrix which is solidified to obtain the final material. Understanding and controlling the cellular organisation of the initially liquid template is therefore of utmost interest in order to control the properties of the cellular solid. This requires the control over the size, organisation and relative volume of the cells, and also over the stability of the liquid template with respect to ageing effects like cell coalescence. While such questions have been investigated successfully for different types of polymers, only few advances have been made for materials with a silicone matrix such as PDMS (polydimethylsiloxane).[1-5] This is due to the difficulty of finding sufficiently efficient stabilizing agents for the liquid template which allow to obtain high volume fractions of the internal phase. In order to tackle this problem, we combine here a reactive blending[6-8] with a polyHIPE (High Internal Phase Emulsion) approach[9-14] using a model system which consists of two immiscible polymers: closely-packed PEG (polyethylene glycol) drops in a continuous phase of siloxane copolymer MHDS (MethylHydrosiloxane - Dimethylsiloxane Copolymers, Trimethylsiloxy terminated) with Si-H groups along the chain. A detailed list of the different molecules and their physical parameters can be found in Table S1 in the supplementary materials. Depending on the emulsion type, the PEG drops occupy between 74 and 99.9% of the volume fraction. These drops are stabilised against coalescence thanks to a crosslinking reaction initiated by a crosslinker/catalyser mixture which is initially dissolved in the PEG droplets. The diffusion of the crosslinker/catalyst molecules to the surface of the droplet creates a solid-like skin of silicone around the PEG drop when both liquids enter into contact. After proper optimisation of the formulation, the created PEG-in-MHDS emulsion is indefinitely stable and the size and organisation of the droplets can be adjusted at will before solidifying the continuous matrix by a second crosslinker/catalyser pair previously dissolved in the continuous phase. Here we show the feasibility of this approach and we discuss in detail the crucial step of ensuring stability of the emulsion template by correlating it with the properties of the PEG/MHDS interface. We show that the optimised formulation can be used to generate macro-porous solids with well-controlled structural properties. The liquid drops are maintained in order to create a "solid emulsion" but may eventually be removed to obtain a porous material.

## 2. Results and Discussion

To stabilize the PEG-in-MHDS emulsion, a vinyl-terminated siloxane crosslinker containing a platinum ($Pt$) catalyst which is active at room temperature, is added in the PEG phase (Table S1) at a concentration $C$ (in mol%). The concentration of catalyst Pt in the crosslinker/catalyst mixture, $F(Pt)$ (in mol%), can be varied between experiments (Experimental Section). Thus the $Pt$ concentration $C_{Pt}$ in the PEG/crosslinker/catalyst mixture depends on the crosslinker/catalyst concentration $C$ in the PEG by the relation: $C_{Pt} = C * F(Pt)$. As the PEG and MHDS are in excess compared to the crosslinker/catalyst, we believe the interfacial properties to depend only on the temperature and the crosslinker/catalyst concentration $C$ in the PEG phase.

To study the emulsion stability, we use a simple protocol: in a cuvette filled with the MHDS, millimetric drops of PEG containing the crosslinker/catalyst mixture (PEG/crosslinker/catalyst mixture) are generated one by one at a constant rate. Being heavier than the MHDS, they pile at the bottom of the cuvette, and the emulsion stability is monitored for different concentrations $C$ (Section S1 in SI). An example is shown in **Figure 1**. We see that there is a critical concentration $C^*$ beyond which the emulsions are indefinitely stable. We interpret this behaviour as the signature of the onset of the formation of a skin-like layer at the interface above $C^*$ as shown on the scheme in **Figure 2a**.

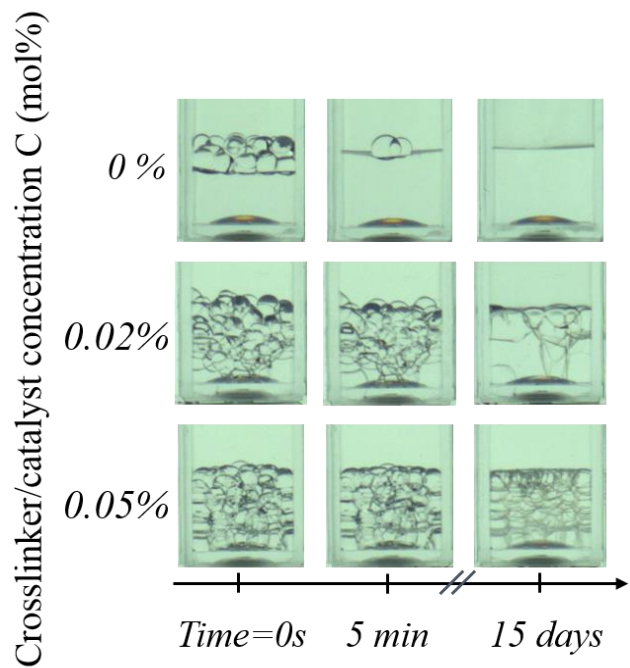

**Figure 1.** Evolution of emulsion stability with time for different concentrations C for the MHDS 2000-25 / PEG400, F(Pt)=0.02 system. The scale bar is 1cm.

To understand the formation of this elastic layer, we need to consider the different competing reactions resulting from the presence of the crosslinker/catalyser mixture (Figure 2b). First it catalyses the cross-linking reaction of the Si-H groups on the MHDS chains in a hydrosilylation reaction with the vinyl ends of the crosslinker (Figure 2b, reaction (1)) which diffuses from the PEG drop into the MHDS phase, thus increasing the molecular mass and finally creating a continuous network of the MHDS chains in the vicinity of the PEG/MHDS interface. However, the Pt also catalyses the oxidation of the MHDS via its Si-H bonds in contact with the PEG phase by the traces of water contained in the PEG, transforming the Si-H in Si-OH bonds while releasing dihydrogen bubbles (Figure 2b, reaction (2)), and preventing the hydrosilylation reaction from happening. [15,16] The observation of bubbles during the emulsion generation at high $C$ is a hint that this reaction effectively happens (Section S2 in SI). The created Si-OH bonds are unstable, and can be transformed in two ways. They have the possibility to react with the C-OH bonds of the PEG to create PDMS-b-PEG copolymers while releasing water (Figure 2b, reaction (3a))[17-19]. These copolymers are also known to be unstable and the inverse reaction would eventually happen. But the Si-OH bonds can also react with each other, leading to a second crosslinking reaction of the MHDS (Figure 2b, reaction (3b)).[20]

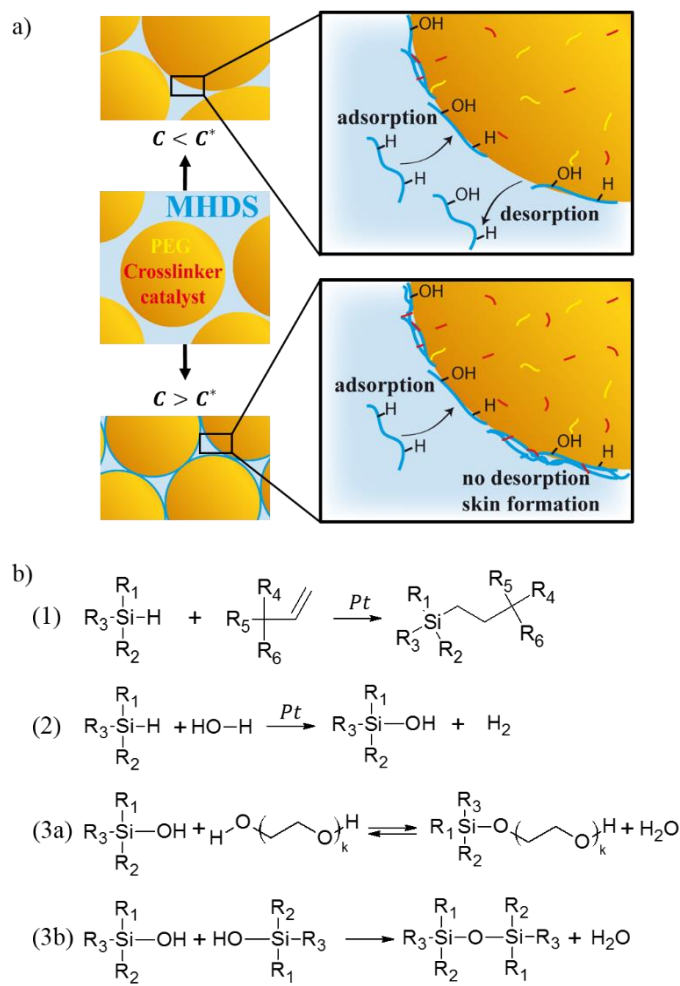

**Figure 2.** a) Scheme representing the stabilisation mechanisms depending on the concentration. Top: regime of adsorption/desorption of MHDS at the interface. Bottom: no desorption of MHDS from the interface, skin formation around the drop. b) Reaction mechanisms occurring at the interface between MHDS and PEG in presence of the crosslinker with its platinum counterpart and traces of water.

We probe the effect of the reactions at the MHDS/PEG interface by measuring the evolution of the interfacial tension with time using a pendant drop of the PEG/crosslinker/catalyst mixture generated in pure MHDS at a constant temperature of 25°C for initial crosslinker/catalyst concentrations $C$ in the PEG phase between 0 and 1 mol% (Experimental section). A decrease in interfacial tension with time is observed in **Figure 3a** until a plateau is reached. Both the characteristic time of the $\gamma(C)$ curves and the plateau value decrease as $C$ increases. We interpret this decrease of interfacial tension as a result of both the reactions (2) and (3a) (Figure 2b). Indeed, reaction (2) creates Si-OH bonds at the interface between the MHDS and the PEG, which have a greater affinity with the PEG than the Si-H bonds. But we believe reaction (3a) to be preponderant in this process, especially since the interfacial tension reaches values which are close to those found in the literature when adding already prepared PDMS-b-PEG copolymers at a PDMS/PEG interface.[21]

We therefore have one set of reactions which is essentially responsible for the reduction in interfacial tension and one which essentially creates the skin-like layer around the drops. However, both phenomena are coupled. In order to better characterize this coupling, the experiment is repeated with variable PEG, MHDS (Table S1) and/or $F(Pt)$. Figure 3b shows the plateau values of the interfacial tension of Figure 3a for all experiment as a function of the platinum concentration $C_{Pt}$. These curves show a non-linear decrease of the plateau value of the interfacial tension with $C_{Pt}$ down to about 1 mN/m at a critical concentration beyond which the interfacial tension is independent of $C_{Pt}$ at long times. We interpret these two regimes in terms of a competition between the kinetics of the different reactions and adsorption/desorption processes which occur at the interface. The catalyst is active at room temperature so we can assume that the kinetics of the reactions at the interface are fast. This leads to the system's global evolution to be governed by two characteristic times (Figure 2a): the diffusion time $\tau_{\text{diff}}$ of the crosslinker/catalyst towards the interface, and the desorption time $\tau_{\text{des}}$ given by the average time a MHDS molecule remains in the vicinity of the interface. The value of $C_{Pt}$ has an immediate effect on $\tau_{\text{diff}}$ only, and different scenarios can then occur. For the lowest values of $C_{Pt}$, the diffusion rate is low, so we can assume $\tau_{\text{diff}} \approx \tau_{\text{des}}$. The interface crosslinking increases slowly the molecular weight of the MHDS chains in the vicinity of the interface, but as for polymers the desorption rate is inversely proportional to their molecular weight, the desorption of these copolymers is still possible, leading to an equilibrium between these two mechanisms.[22] The plateau value of interfacial tension depends then on $C_{Pt}$ during this adsorption/desorption regime (Figure 2a top). Inversely, for the highest values of $C_{Pt}$, we can assume that $\tau_{diff} > \tau_{des}$, meaning that the MHDS at the interface will be linked to the other chains close to it to form a molecule of higher molecular weight before it can desorb from the interface. In this regime (Figure 2a bottom), the MHDS will not desorb and a skin-like layer will form at the interface, leading to the independence of the plateau value of interfacial tension on $C_{Pt}$.

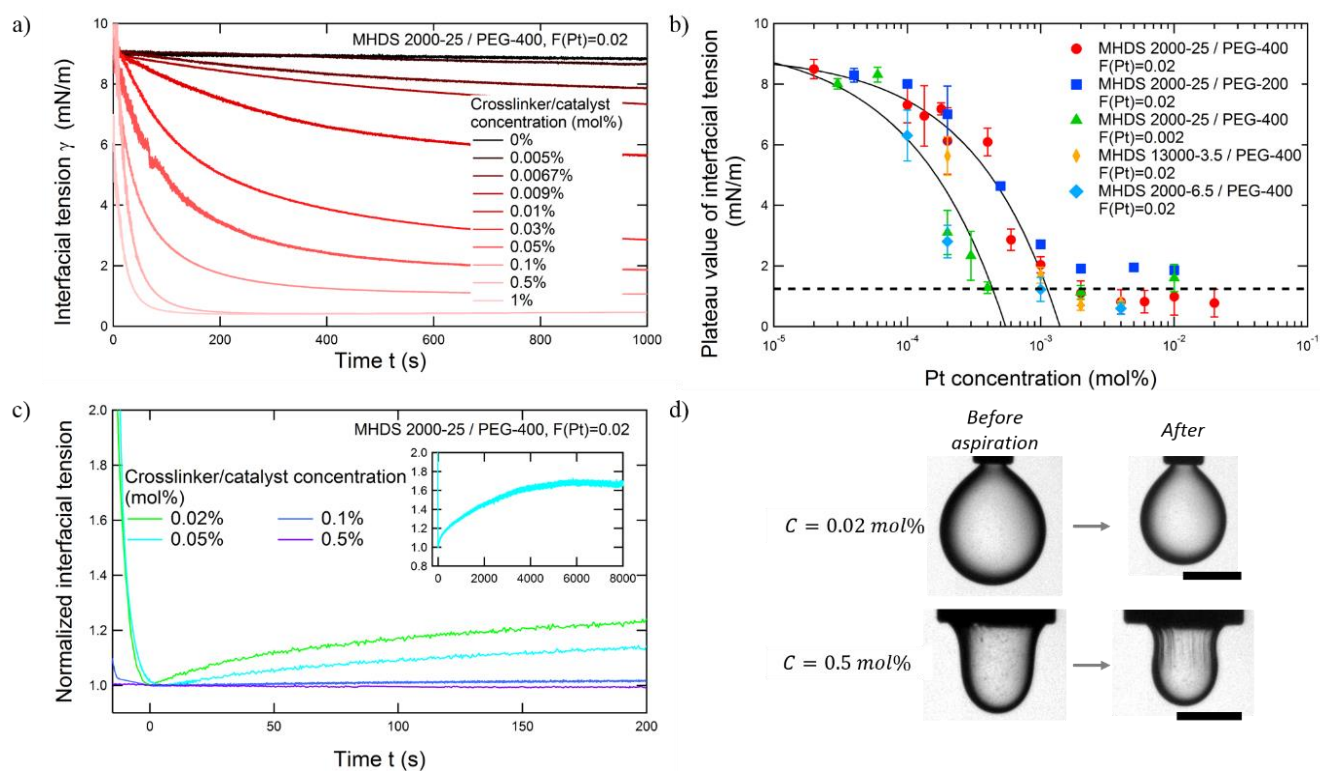

**Figure 3.** a) Time evolution of the interfacial tension between MHDS 2000-25 and PEG400 with F(Pt) = 0.02 for different concentrations of crosslinker/catalyst in the PEG phase. b) Plateau value of the interfacial tension between MHDS and PEG as a function of catalyst concentration in the PEG phase. The different colours and markers corresponds to different systems with different oligomer sizes, F(Pt), or number of reactive sites along the MHDS backbone, which are summarised in Table S1. The solid lines are guide to the eye. c) Time evolution of the normalized interfacial tension after a rapid volume reduction of the drop at t=0s, for the MHDS 2000-25/PEG400, F(Pt) = 0.02 system. The insert is the long-time evolution for $C = 0.05$ mol%. d) Pictures of the experiment c) before and after volume reduction of the drop for two crosslinker/catalyst concentrations in the two different regimes identified on b) with their respective scale bars of 1 mm.

These data also suggest that the plateau value of the interfacial tension seems to be independent of the MHDS and PEG sizes in this molecular weight range, which is coherent with our hypothesis that the interfacial tension evolution is the result of the reactions (2) and (3a) only. However the value of $F(Pt)$ and the number of Si-H bonds along the MHDS backbone change the critical concentration at which the plateau value of interfacial tension is independent of $C_{Pt}$. Since both of these parameters have an immediate impact on the reactions kinetics, this remains coherent within our hypothesis (Section S3 in SI).[19]

To test the desorption hypothesis, we create a drop of the PEG/crosslinker/catalyst mixture in the MHDS and leave it at rest for a sufficiently long time for the system to reach its plateau value for different crosslinker/catalyst concentrations $C$. The volume of the drop is then abruptly reduced by aspiration of liquid and the evolution of the interfacial tension is followed (Figure 3c) while the drop volume is kept constant at its new value.[23] Upon rapid volume reduction, the interfacial tension also decreases since the hydrophilic groups on the MHDS surface are compacted. For low crosslinker/catalyst concentrations ($C < C^*$), the surface tension increases again until it reaches a final constant value, lower than its initial plateau value (insert Figure 3c). We believe this relaxation to be due to the desorption of the HO-MHDS or PDMS-b-PEG copolymers from the interface. For high concentrations, the surface tension remains constant after volume reduction, indicating that the molecules cannot desorb from the MHDS/PEG interface. It that case, wrinkles are also seen at the interface (Figure 3d), which is the signature of an elastic skin. The crosslinker/catalyst concentration at which the relaxation stops correlates well with the results seen Figure 1 and Figure 3b. This means that the onset of the plateau in Figure 3b - and hence $C^*$ - can be associated with the creation of a continuous skin around the drop as assumed earlier.

We therefore summarise here our hypothesis that for long times the critical crosslinker/catalyst concentration $C^*$ marks the separation of two regimes. At $C < C^*$ the reactions at the interface lead to a reduction in interfacial tension and to a global increase in molecular mass of the MHDS, but the system remains liquid-like. On the contrary, at $C > C^*$, the crosslinking reaction leads to the formation of an elastic skin around the droplet which also correlates with the formation of stable emulsions (Figure 1 and 3b).

Based on the above discussion we can optimise the formulation, taking into account that $C \geq C^*$ to ensure skin formation. For our standard formulation we therefore use $C = 0.05$ mol% with $F(Pt) = 0.02$.

Since no surfactant was added initially to stabilize the emulsions, PEG-in-MHDS and MHDS-in-PEG emulsion configurations are equally probable, so classical emulsification techniques such as turbulent mixing lack a selection mechanism and generate mixed emulsions (Section S4 in SI). In order to tackle this, and to generate emulsions with controllable drop sizes, we use alternative techniques which explicitly select the dispersed phase (Experimental section): millimetric droplets are generated by dispensing the PEG/Crosslinker/catalyser mixture from a syringe at constant flow rate using a syringe pump (**Figure 4a**), whereas smaller drops (200 – 500 µm) are generated by millifluidic techniques such as a flow-focusing device or a T-junction device (Figure 3b).[24,25]

By varying the flow rates and/or the geometric dimensions of the different devices, we are able to generate stable emulsions with a wide range of volume fractions and with a wide range of drop radii (100 µm up to several millimeters depending on the flow rates). Varying the aspect ratio of the drop-to-container size we can generate ordered (Figure 4c) or disordered emulsions (Figure 4d to j). Moreover, by varying the volume fraction of the continuous phase the drop geometry can be varied from polyhedral (Figure 4c, d and e) to nearly spherical (Figure 4f to i).

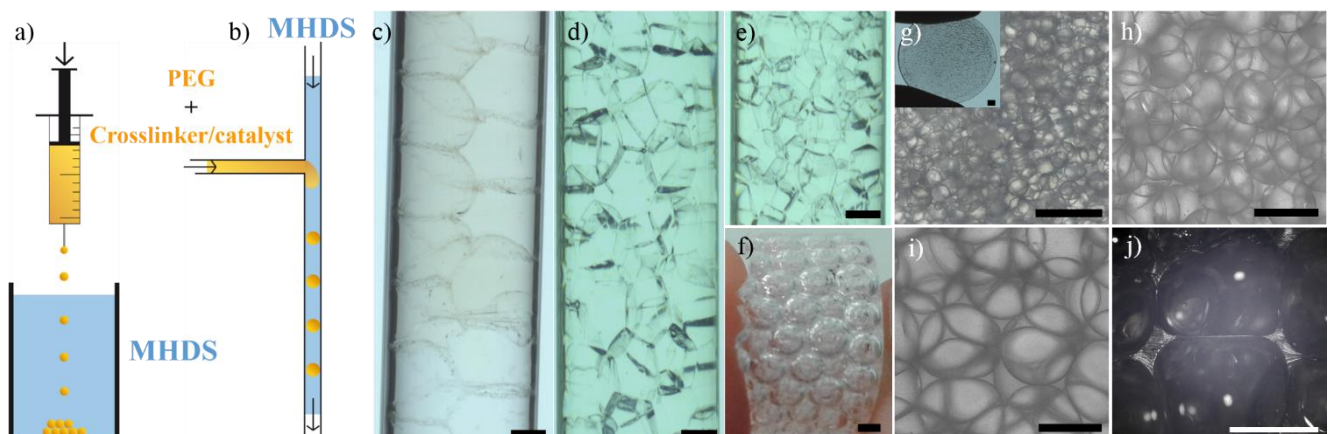

**Figure 4.** a) Scheme of the generation of millimetre-sized PEG/crosslinker/catalyser drops in MHDS by simple dripping from a needle using a syringe pump. b) Scheme of the generation of drops with diameters of the order of 1mm with a T-junction device. c), d) and e) Liquid PEG-in-MHDS emulsions with high drop content and tube diameter over drop diameter ratio of 2, 3 and 4.6 respectively, generated with MHDS 2000-25 and PEG400. f) Solid silicone foam made from a water-in-Sylgard® 184 template, g), h) and i) Solid emulsions generated with Sylgard® 184 and PEG400, with drop radius of respectively 335 µm, 850 µm and 1390 µm. The insert is a picture of the solid emulsion g). j) Close-up image of the interface between two drops in a PEG-in- Sylgard® 184 solid emulsion. All scale bars are 2 mm.

Finally, the solidification of the MHDS phase was done by the addition of another platinum-based crosslinker (platinum cyclovinylmethylsiloxane complex Table S1) in the silicon phase prior to emulsification, which is active at moderate temperatures and which we call here the "solidifier". Even if the same hydrosilylation reaction is responsible for the crosslinking of the MHDS/PEG skin-like layer and the solidification of the MHDS matrix, we believe that the addition of the solidifier does not interfere with the different chemical processes discussed above. Indeed, the liquid emulsions – and the stabilising layer - are generated at ambient temperature and at timescale which are shorter than the solidifier reaction time at this temperature. We therefore generate the liquid emulsion at room temperature with the same interfacial properties as discussed above. These are then cured for a few hours at 50°C in oven. The presence of the skin-like layer ensures that no coalescence occurs between the drops during the curing stage. By solidifying the MHDS phase, one obtains a macro-cellular elastomer, which we call a "solid emulsion" (Figure 4g to j) composed of a PDMS matrix with liquid PEG inclusions. Figure 4j is a photo of a cut of one sample showing the zone of contact between two drops in a solid PEG-in- Sylgard® 184 emulsion after removing the PEG liquid phase. The mean thickness of the PDMS film between two drops of PEG is approximately 25µm in the center of the film and is reduced when approaching the plateau border.

We have also been able to create silicone foams using the same approach, by replacing the PEG phase by water (in which the crosslinker/catalyst mixture is dissolved). After evaporation of the water an open-cell silicone foam (Figure 4f and section S5 in SI) is obtained whose structure can be tuned in the same way as the emulsions.

## 3. Conclusions

In conclusion, we have reported for the first time the generation of ultra-stable emulsion with PDMS as continuous phase, via reactive blending as a stabilisation route, with drop sizes up to the millimetre scale. We believe this reactive blending approach to be transferable to other polymeric systems. For example, the in-situ formation of block-copolymers at droplet surfaces is routinely used for a wide range of polymer blends [6-8]. And the formation of a skin on the droplet by initiating the polymerisation from the dispersed phase is used for the generation of polyHIPEs with different polymer systems (yet with the additional use of a surfactant to avoid droplet coalescence)[26,27]. Our work combines these two aspects. The obtained two-phase PDMS materials, either in the liquid state or solidified, present original structures as compared to hard granular materials or soft materials such as foams, because of their deformable but frictional interface.[28,29] Moreover, the presence of the liquid drops inside a purely elastic matrix modifies greatly the mechanical properties of the material as compared to the ones of the bare PDMS matrix. The size, organisation and volume fraction of the drops can be used to fine-tune the mechanical properties of the solid emulsion. Future work will concentrate on the control of the emulsion structure and its relationship with its visco-elastic properties.

## 4. Experimental Section

*Interfacial study:* All the molecules Table S1 are used as received. The crosslinker/catalyst concentration is in mol% in the PEG, and the mixtures are made in clean bottles. The $F(Pt)$ is originally 0.02, and can be reduced by diluting the original mixture with the same siloxane chains (pure crosslinker).

The interfacial measurements are made using a pendant drop apparatus (Tracker from Teclis), using the Laplacian profile method with a regulation of temperature.[30] The PEG drops are generated in MHDS at a constant velocity and their volume ranges between 2 and 10µL.

*Optical images*: The images Figure 1 and Figure 4c to f were made using a uEye camera. The close-up optical images of the solid emulsions Figure 4g to i were taken with a Keyence microscope (VHX-2000).

*Millifluidic devices*: The flow-focusing devices were designed on the SolidWorks software and drilled on PMMA (Poly(methyl methacrylate)) using a micro milling machine. The constriction are 100 µm in width and 500 µm in length and height. A PDMS casting mold was made from these and used to create COC (Cyclic olefin copolymer) chips with a hot press at 130°C. This chip is closed with a crosslinked PDMS sheet.

The T-Junctions are commercial ones from Nordson Medical (T10-6005).

## Supporting Information

Supporting Information is available from the Wiley Online Library or from the author.

Acknowledgements: We acknowledge funding from the European Research Council (ERC) under the European Union's Seventh Framework Program (FP7/2007-2013) in form of an ERC Starting Grant, agreement 307280-POMCAPS.

The authors would like to thank Anniina Salonen, Emmanuelle Rio, Dominique Langevin, Mehdi Zeghal and Gero Decher for fruitful discussions.

Received: May 13, 2016; Revised: June 22, 2016; Published online: July 28, 2016; DOI: 10.1002/marc.201600281

Keywords: Poly-HIPE, Solid emulsions, Reactive blending, Interfacial tension, Cellular polymers

**Macrocellular silicone polymers** are obtained by solidification of the continuous phase of a PDMS (polydimethylsiloxane) emulsion, which contains PEG (polyethyleneglycol) drops of sub-millimetric dimensions. Coalescence of the liquid template emulsion is prohibited by a reactive blending approach, which could serve as a model emulsification route for any type of polymeric systems.

A. Giustiniani*, P. Guégan, M. Marchand, C. Poulard and W. Drenckhan

**Generation of silicone poly-HIPEs with controlled pore sizes via reactive emulsion stabilization**

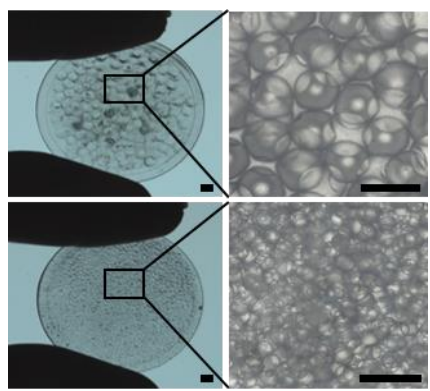

## Supporting Information

**Generation of silicone poly-HIPEs with controlled pore sizes via reactive emulsion stabilization**

Anaïs Giustiniani*, Philippe Guégan, Manon Marchand, Christophe Poulard and Wiebke Drenckhan

**Table S1.** Molecules used for this study and their relevant properties

| Molecule | Name used | $M_w$ [g.mol$^{-1}$][a] | $\rho$ [g.mL$^{-1}$][b] | Characteristics | Supplier |
|---|---|---|---|---|---|
| MethylHydrosiloxane - Dimethylsiloxane Copolymers, Trimethylsiloxy terminated (MHDS) | MHDS 2000-6.5<br>MHDS 2000-25<br>MHDS 13000-3.5 | 2000<br>2000<br>13000 | 0.97<br>0.98<br>0.97 | Mole%(MeHSiO)=6.5%<br>Mole%(MeHSiO)=25%<br>Mole%(MeHSiO)=3.5% | Gelest<br>Gelest<br>Petrarch systems |
| Polyethyleneglycol (PEG) | PEG200<br>PEG400 | 200<br>400 | 1.124<br>1.128 | | Sigma Aldrich<br>Sigma Aldrich |
| Platinum(0)-1,3-divinyl-1,1,3,3-tetramethyldisiloxane complex solution 0.1 M in poly(dimethylsiloxane), vinyl terminated | Crosslinker/catalyst | 381.48 | 0.98 | Amb. Temp. | Gelest |
| Platinum-cyclovinylmethylsiloxane complex | Solidifier | 539.74 | 1.02 | Mod. Temp. | Gelest |
| | Sylgard® 184 | | 1.03 | Ratio MHDS:crosslinker = 10:1<br>$\eta_{MHDS} = 5000$ cP<br>$\eta_{MHDS+crosslinker} = 3500$ cP | Dow Corning |

[a] Molecular weight of the molecule; [b] Density at 25°C

**S1: PEG-in-MHDS emulsion stability**

To study the emulsion stability, we use a simple protocol: in a cuvette filled with the MHDS, millimetric drops of the PEG containing the crosslinker/catalyst mixture (PEG/crosslinker/catalyst mixture) are generated one by one at a 0.1 mL.min$^{-1}$ rate with a syringe pump (World Precision Instrument, AL-1000). The emulsion stability is monitored for different crosslinker/catalyst concentrations by taking one image every 45 s at the beginning of the experiment when the emulsion is evolving rapidly, and one every 1 hour after, using a digital camera (u-eye camera). An example of a sequence of photographs for some of the tested concentrations with the definition of the emulsion height is shown in **Figure S1a** for the system MHDS 2000-25 / PEG400, $F(Pt) = 0.02$. The emulsion generation is prior to the beginning of the imaging at time $t = 0\ s$, so the $h_0$ is different for every concentration since the emulsion has already started evolving. Figure S1b shows the graphical representation of the evolution of the normalized emulsion height for every $C$. We see that there is a critical concentration $0.03\ mol\% \leq C^* \leq 0.05\ mol\%$ beyond which the emulsions are stable indefinitely.

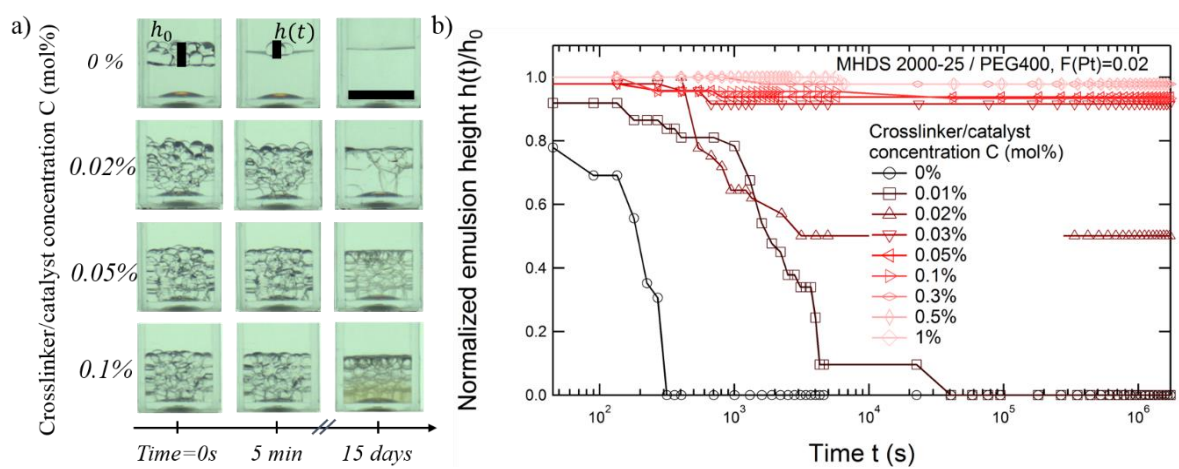

**Figure S1.** Evolution of emulsion stability with time for different concentrations $C$ for the MHDS 2000-25 / PEG400, $F(Pt) = 0.02$ system. a) Photographs of emulsions at different times showing the definition of $h(t)$ and $h_0$ as the emulsion height at time $t$ and $t = 0\ s$ respectively. The scale bar is 1cm. b) Graphical representation of the evolution of the normalized emulsion height $h(t)/h_0$ with time.

A brown coloration appears on the emulsions for high crosslinker/catalyst concentrations ($C \geq 0.1\ mol\%$), and its intensity increases with $C$. This is due to the presence of $Pt$ nanoparticles forming after consumption of the crosslinkers, and can be avoided by choosing $C = 0.05\ mol\%$ for the final material.

**S2: Generation of visible H₂ bubbles at high crosslinker/catalyst concentration**

The reactions taken into consideration to understand the evolution of interfacial tension between the MHDS and the PEG are shown Figure 2c. Reaction (2) releases H$_2$ bubbles. The observation of bubbles during the emulsion generation (**Figure S2**) is a hint that this reaction effectively happens. These H$_2$ bubbles can stay trapped in the emulsion, thus causing imperfections in the emulsion structure. The size of these bubbles correlates with the crosslinker/catalyst concentration $C$, and visible bubbles are mostly seen for $C > 0.1\ mol\%$ only. By keeping $C = 0.05\ mol\%$ for the emulsion generation, we have been able to avoid any bubble formation.

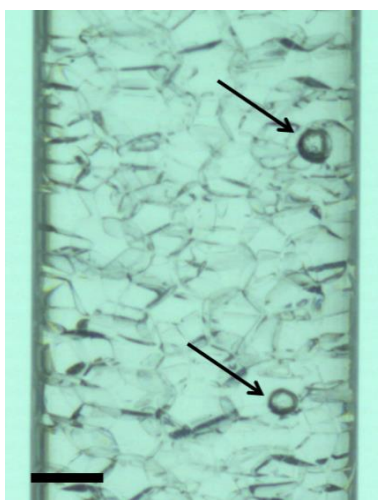

**Figure S2.** Photograph of a PEG-in-PDMS emulsion with H$_2$ bubbles indicated by arrows. The scale bar is 2 mm.

**S3: Interfacial tension between MHDS and PEG**

From the data shown in Figure 3b, we observe that the plateau value of interfacial tension seems to be independent of the molecular weight of the MHDS and the PEG, as all the concerned data follow a single master curve (red, orange and dark blue markers). The reader should note here that the number of reactive sites per chain on the MHDS 2000-25 and 13000-3.5 are approximately equal. When decreasing $F(Pt)$ by a factor 10 (green markers in Figure 3b), the final plateau value stays the same within the error bars, but we lower the concentration $C^*$ at which the plateau is reached. Upon a decrease of $F(Pt)$ we promote the hydrosilylation over the oxidation reaction.[16] For the same amount of $Pt$ in the two MHDS 2000-25 / PEG400, $F(Pt) = 0.02$ and $F(Pt) = 0.002$ systems, the crosslinking of the interface is therefore more efficient in the latter case, and consequently the plateau value is reached at a lower $C_{Pt}$ since the crosslinking will prevent the Si-OH bonds or the PDMS-b-PEG copolymers to desorb from the interface or new ones to be added. The same observation is made when the number of Si-H bonds on the MHDS backbone changes with a fixed MHDS size (light blue markers Figure 3b), the creation of the skin occurs at a lower concentration. In this case, the different reaction kinetics are the same since the ratio $F(Pt)$ is constant, meaning that for a given time, there will be as many sites that have reacted for MHDS 2000-25 and MHDS 2000-6.5, and we can then consider only the hydrosilylation reaction for this interpretation. As there are less reactive sites on MHDS 2000-6.5, less crosslinker molecules are needed to connect all the MHDS chains at the interface, thus the lower $C^*$ value. This also means that these sites will react with other sites on chains that are farther away than in the case of MHDS 2000-25.

**S4: Mixed emulsions**

The classic emulsification techniques, such as turbulent mixing, lack a selection mechanism and generate mixed emulsions with our systems. **Figure S3** shows an optical image of a mixed emulsion of PEG and MHDS obtained with a confocal microscope (Leica TCS SP8) with fluorescein, a fluorescent dye, added in the PEG phase. In this image, one can see PEG drops in MHDS (green drops on a black background) and MHDS drops in PEG (black drops on a green background).

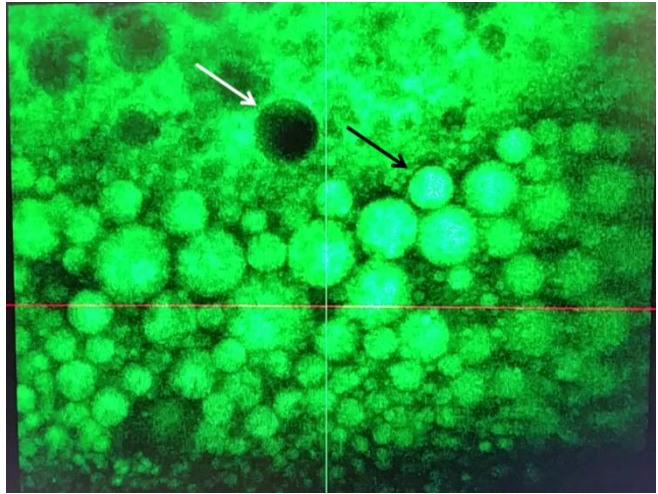

**Figure S3.** Optical image obtained with a confocal microscope of an emulsion generated by turbulent mixing of PEG and MHDS. The black arrow points towards a drop of PEG surrounded by MHDS. The white arrow points towards an MHDS drop surrounded by PEG. Fluorescein was added in the PEG phase. The average drop size is of the order of 10 µm.

**S5: Silicone foam**

By replacing the PEG phase by water (in which the crosslinker/catalyst mixture is dissolved), and after evaporation of the water, a silicone foam (Figure 4f) is obtained whose structure can be tuned in the same way as the emulsions. Looking at the silicone foams through a microscope shows it to have an open-cell structure, as seen **Figure S4**.

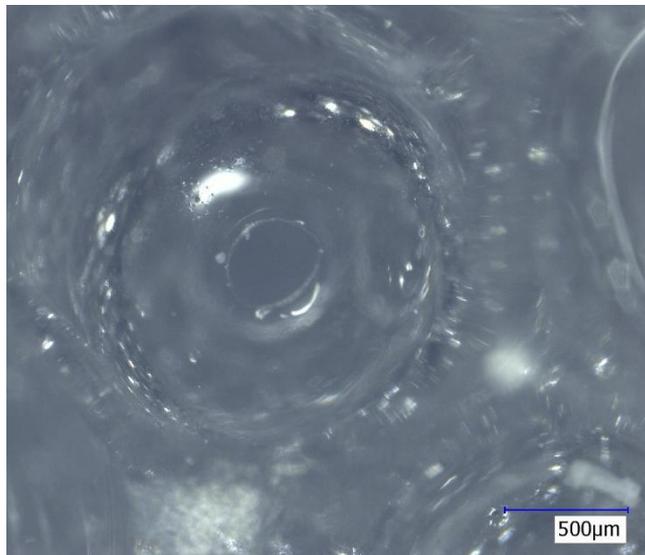

**Figure S4.** Optical image obtained with a microscope of an open pore of a silicone foam generated by emulsion templating.